\documentclass[epj-spec]{svjour}
\usepackage{graphicx}
\usepackage{bm}
\usepackage{amssymb}
\begin{document}
\title{Jozs\'o's Legacy: Chemical and Kinetic Freeze-out in Heavy-Ion 
Collisions}
\author{Ulrich Heinz\inst{1,2}\fnmsep\thanks{\email{heinz@mps.ohio-state.edu}. 
Work supported by the U.S. Department of 
Energy, grant DE-FG02-01ER41190.}
\and Gregory Kestin\inst{1}\fnmsep\thanks{Work supported by the U.S. National 
Science Foundation, grant PHY-0354916.}}
\institute{
Department of Physics, The Ohio State University, Columbus, OH 43210, USA
\and 
Physics Department, Theory Division, CERN, CH-1211 Geneva 23, Switzerland}
\abstract{We review J.\,Zim\'anyi's key contributions to the theoretical
understanding of dynamical freeze-out in nuclear collisions and their
subsequent applications to ultra-relativistic heavy-ion collisions,
leading to the discovery of a freeze-out hierarchy where chemical 
freeze-out of hadron yields precedes the thermal decoup\-ling of their
momentum spectra. Following Zim\'anyi's lines of reasoning we show that 
kinetic freeze-out necessarily leads to a dependence of the corresponding
freeze-out temperature on collision centrality. This centrality 
dependence can be predicted within hydrodynamic models, and for Au+Au
collisions at RHIC this prediction is shown to reproduce the experimentally
observed centrality dependence of the thermal decoupling temperature,
extracted from hadron momentum spectra. The fact that no such centrality
dependence is observed for the chemical decoupling temperature, extracted
from the hadron yields measured in these collisions, excludes a similar
kinetic interpretation of the chemical decoupling process. We argue that
the chemical decoupling data from Au+Au collisions at RHIC can only be 
consistently understood if the chemical freeze-out process is driven by
a phase transition, and that the measured chemical decoupling temperature
therefore measures the critical temperature of the quark-hadron 
phase transition. We propose additional experiments to further test
this interpretation.  
} %end of abstract
\maketitle

%
%%%%%%%%%%%%%%%%%%%%%%%%%%%%%%%%%%%%%%%%%%%%%%%%%%%%%%%%%%%%%%%%%%%%%%%%%%%
\section{Jozs\'o's pioneering work and early encounters}
\label{intro}
%%%%%%%%%%%%%%%%%%%%%%%%%%%%%%%%%%%%%%%%%%%%%%%%%%%%%%%%%%%%%%%%%%%%%%%%%%%
%
Due to political complications, J\'ozsef Zim\'anyi began his long and 
successful scientific career as an experimental physicist. 
However, in 1977 Jozs\'o jump-started his reputation as an outstanding
nuclear theorist by writing, together with J.\,P.\,Bondorf and 
S.\,I.\,A.\,Garpman, a very influential paper with the title ``A simple 
analytical model for expanding fireballs'' \cite{BGZ78}. In this paper,
Jozs\'o and his friends discovered a class of scaling solutions of the 
non-relativistic Euler equations for the hydrodynamic evolution of 
spherically symmetric fireballs, with power-law radial density and 
velocity profiles, which, due to its symplicity and elegance, has
continued until today to spawn follow-up papers (in particular by his 
students and colleagues T.\,Bir\'o and T.\,Cs\"org\H o) generalizing it 
to systems with less symmetry and undergoing relativistic expansion. 

More important for the present work is, however, Section 3 of that
paper \cite{BGZ78}, with the title ``Geometric concept of the break-up''.
This section heading is slightly misleading since what is being developed
is really a {\em dynamic} freeze-out concept (even if it is derived with
the help of a geometric sketch of the expanding fireball). It introduces
the ideas of the competition between (i) the ``separation velocity'' 
of neighboring volume elements (in modern language: flow velocity gradients) 
and the thermal velocity, and (ii) the expansion rate (again controlled
by velocity gradients) and the local scattering rate, as drivers of the 
freeze-out process. A key point following from these ideas, although
not emphasized in \cite{BGZ78}, is that the thermal velocity
and scattering rate both depend on the particle species, leading to
the concept of {\em differential freeze-out} where different particles
(or processes, see below) decouple at different times.   

The space-time trajectories of Zim\'anyi and one of us (UH) first joined 
for extended periods in the summers of 1985/86 when UH, employed by 
Brookhaven National Laboratory and preparing for the ``imminent'' start 
of the Relativistic Heavy-Ion Collider (RHIC eventually became operational 
in 2000!), worked together with Mark Rhoades-Brown at the University of 
Stony Brook and our common student Kang Seok Lee on the problem of 
hadronization of a quark-gluon plasma (QGP), while Jozs\'o (supported by an 
NSF-Hungarian Academy of Sciences exchange program) visited his friend 
N\'andor Bal\'azs at Stony Brook to work on the same problem. We discussed 
a number of thermodynamic issues related to this hadronization process, 
having to do with subtleties of the Maxwell construction between a QGP 
and a thermally and chemically equilibrated hadron resonance gas (HG) 
when two different types of quantum numbers (baryon number and strangeness) 
needed to be conserved, without really solving some of the questions that 
flummoxed us before our paths separated again. In fact, our group first got 
it wrong in a paper that we wrote early in 1986 \cite{Lee:1986mn} and 
whose puzzling (and, as it turned out, incorrect) results spurred 
Jozs\'o to look more closely into the problem. In early 1987 three 
papers appeared in short sequence \cite{LZB87,GKS87,HLR87} which 
independently (although clearly triggered by discussions between 
members of the different author groups -- Horst St\"ocker visited
Stony Brook and BNL in 1986, too) solved the problem correctly (Jozs\'o 
\cite{LZB87} beat us all to it by a few months). These papers also 
pointed out the possibility of strangeness separation in a first-order 
QGP-HG phase transition. For those who experienced Jozs\'o's knowledge 
and love of good wines and spirits it should not be too surprising that 
he was the first to mention the analogy of this process with that of 
distilling alcohol, and to coin the phrase ``strangeness distillation'' 
\cite{fn1}.   

Mark, Kang Seok and UH tried to find a way to experimentally measure
this ``distillation'' effect, picking up an idea of Shoji Nagamiya
\cite{N81} who exploited the different mean free paths of $K^+$ mesons 
and protons and pions in nuclear matter to explain the observed hierarchy 
$T_{K^+}>T_p>T_\pi$ of the slopes of their energy spectra measured in 
heavy-ion collisions at the {\tt BEVALAC}. We wanted to apply it to 
$K^+$ and $K^-$ mesons which carry opposite strangeness. While trying to 
understand the influence of collective expansion of the fireball and its
interplay with the mean free path in the freeze-out process, we ran into 
Jozs\'o Zim\'anyi who once again visited Stony Brook. That's when he 
pointed us to his 1978 paper with Bondorf and Garpman \cite{BGZ78}. The 
result of this interaction was a paper on ``$K^+$ and $K^-$ slope 
parameters as a signature for deconfinement at finite baryon density'' 
\cite{Heinz:1987ca} which discusses (we believe) for the first time the 
interplay of differential freeze-out and radial flow, using Jozs\'o's 
hydrodynamic model from 1978 \cite{BGZ78} to calculate the expansion rate 
in the dynamical freeze-out criterium $\tau_\mathrm{exp}=\frac{1}{\partial
\cdot u} < \tau_\mathrm{scatt}=\frac{1}{\rho\langle\sigma v\rangle}$, 
motivated by that same 1978 paper. Since the cross section $\sigma$ and 
hence the scattering rate $\tau_\mathrm{scatt}$ depends on the particle 
species, and the density of scatterers $\rho$ depends on the temperature 
and baryon chemical potential, this freeze-out criterium leads to 
different freeze-out points $(T_f,\mu_f)$ for $K^+$ and $K^-$, with 
$K^+$ predicted to freeze out earlier, and thus with less radial flow, 
than $K^-$. Due to the strangeness distillation effect and the reheating 
of the matter during a first-order phase transition at finite net baryon 
density, we predicted a change of the ordering between the $K^+$ and 
$K^-$ slopes with and without a quark-hadron phase transition 
\cite{Heinz:1987ca}. This works only in fireballs with large net baryon 
density, however, since in baryon-free matter the difference between the 
mean free paths of $K^+$ and $K^-$ disappears.

%
%%%%%%%%%%%%%%%%%%%%%%%%%%%%%%%%%%%%%%%%%%%%%%%%%%%%%%%%%%%%%%%%%%%%%%%%%%%
\section{Dynamic freeze-out and the chemical-thermal decoupling hierarchy}
\label{sec2}
\subsection{Dynamic freeze-out at approximately constant decoupling
temperature}\label{sec2.1}
%%%%%%%%%%%%%%%%%%%%%%%%%%%%%%%%%%%%%%%%%%%%%%%%%%%%%%%%%%%%%%%%%%%%%%%%%%%
%
In 1987 UH moved to the University of Regensburg. One of his first three
students there, Ekkard Schnedermann, further developed the expanding 
fireball description of hadron spectra from heavy-ion collisions in his
diploma thesis (1989, unpublished) and generalized it to a genuinely
dynamical ``global hydrodynamics'' model (with azimuthal instead of
spherical symmetry, assuming longitudinal boost-invariance) in his PhD
thesis (1992) \cite{Schnedermann:1992hp}. By applying Jozs\'o's kinetic 
freeze-out criterium as written above, and also comparing it with a
geometric picture where freeze-out happens once the mean free path 
exceeds the fireball radius, he was able to demonstrate two important 
facts: 
%
%%%%%%%%%%%%%%%%%%%%%% Fig. 1 %%%%%%%%%%%%%%%%%%%%%%%%%%%%%%%%%%%%%%%%%%%
\begin{figure}[htb]
\includegraphics*[width=0.5\columnwidth,height=6cm]{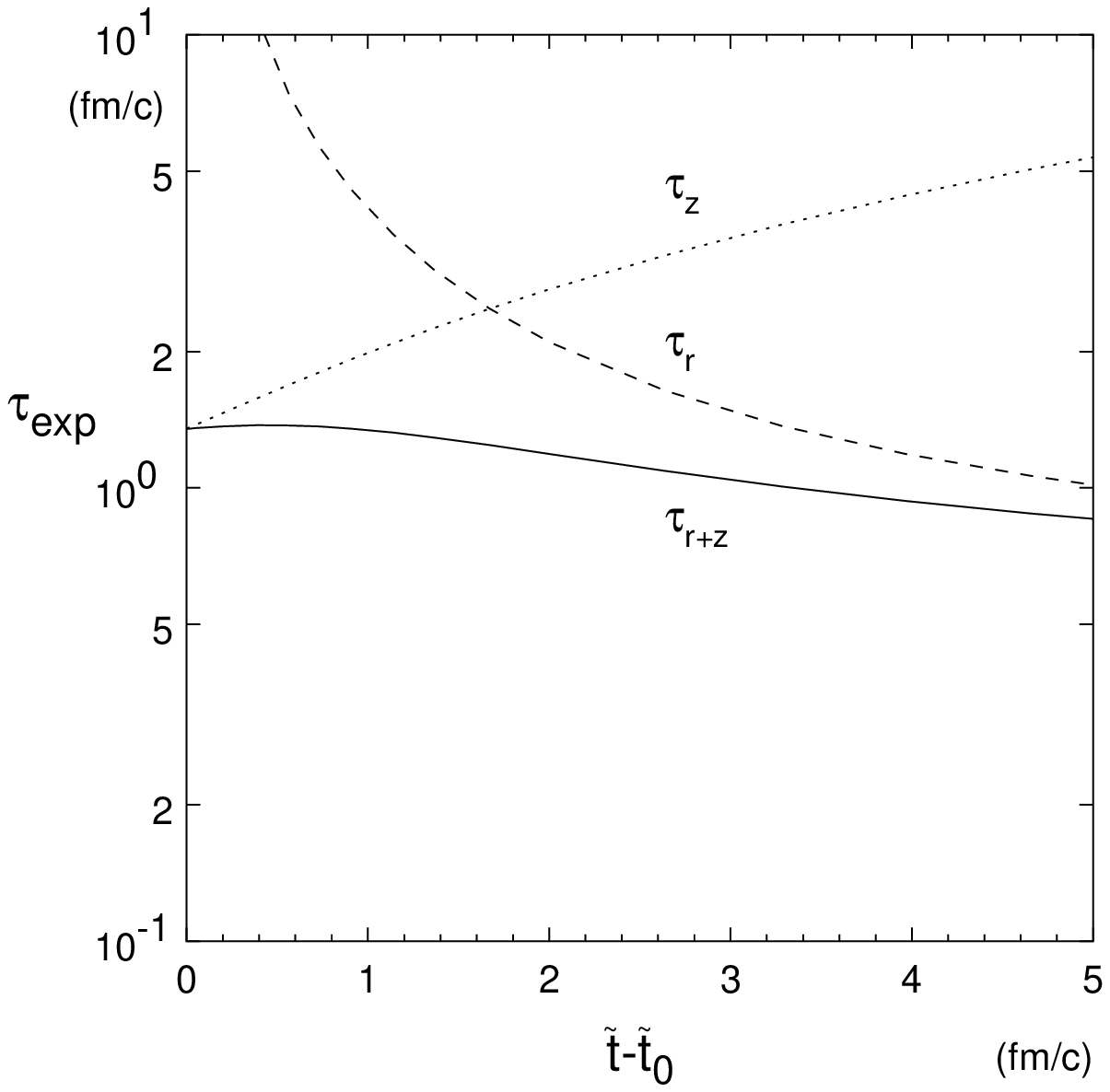}
\includegraphics*[width=0.5\columnwidth,height=6cm]{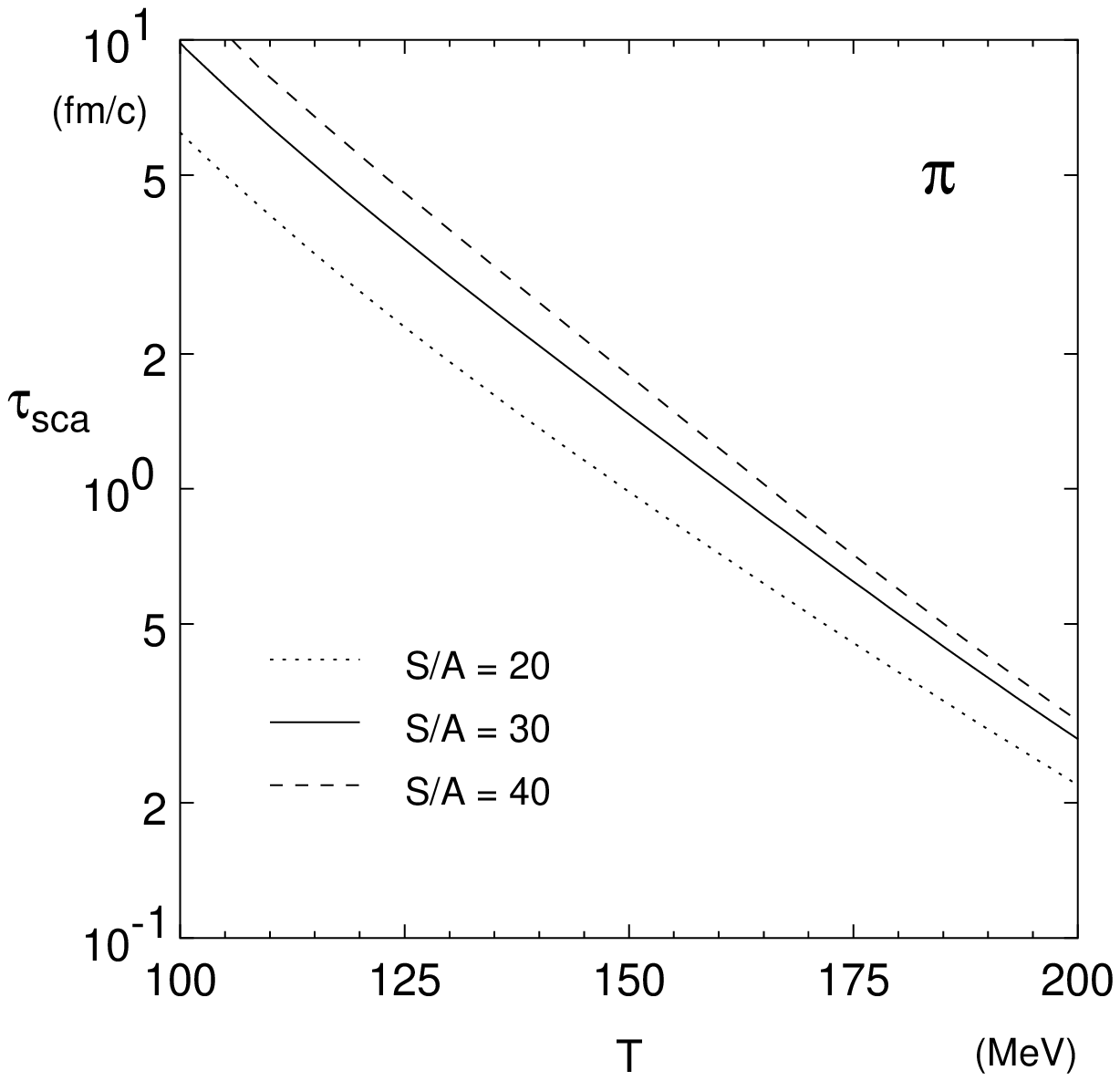}
\caption{Expansion (left panel) and scattering time scales (right panel, 
         for pions in hot nuclear matter with different values of entropy
         per baryon $S/A$ as indicated) for azimuthally expanding fireballs
         formed in S+S collisions at the SPS. For the expansion time
         scale, contributions from longitudinal ($\tau_z$) and
         transverse expansion ($\tau_r$) are shown separately. Figures 
         taken from the second paper in \cite{Schnedermann:1992hp}.}
\label{F1}
\end{figure}
%%%%%%%%%%%%%%%%%%%%%%%%%%%%%%%%%%%%%%%%%%%%%%%%%%%%%%%%%%%%%%%%%%%%%%%%%
%  
(i) In relativistic heavy-ion collisions, freeze-out is driven
by the collective expansion. In particular due to the accelerating 
transverse flow (which compensates for the slowing boost-invariant 
longitudinal expansion, see left panel in Fig.~\ref{F1}), the ``Hubble 
radius'' $\sim\tau_\mathrm{exp}$ of the exploding fireball in the 
``Little Bang'' is always much smaller ($\sim 1\,{-}\,2$\,fm/$c$) than
its geometric radius ($\sim 4$\,fm in S+S), so freeze-out is controlled
by dynamics, not by geometry, just as in the Big Bang. (ii) Due to its 
dependence on the density of scatterers $\rho$, the scattering time scale 
$\tau_\mathrm{scatt}$ has a very steep (exponential) temperature 
dependence (right panel in Fig.~\ref{F1}). Freeze-out (i.e. equality
of the scattering and expansion time scales) thus happens approximately
at constant temperature $T_\mathrm{dec}\approx \mathrm{const}$.

%%%%%%%%%%%%%%%%%%%%%%%%%%%%%%%%%%%%%%%%%%%%%%%%%%%%%%%%%%%%%%%%%%%%%%%%%%%%
\subsection{Differential chemical and thermal freeze-out}\label{sec2.2}
%%%%%%%%%%%%%%%%%%%%%%%%%%%%%%%%%%%%%%%%%%%%%%%%%%%%%%%%%%%%%%%%%%%%%%%%%%%%
%
At this time we also began to realize that there is a conceptual 
difference between chemical and thermal freeze-out in heavy-ion collisions.
\textbf{\emph{Chemical freeze-out}} describes the point where 
{\em inelastic processes} that convert one kind of hadronic species into 
a different one cease and the hadron abundances stop changing. 
\textbf{\emph{Thermal freeze-out}} defines the point where the momenta 
of the particles stop changing, i.e. where {\em all types of 
momentum-changing collisions, elastic and inelastic} cease.
Zim\'anyi's kinetic freeze-out criterium predicts that in a dynamically
evolving fireball these two freeze-out points do not coincide. To see
how this comes about let us rewrite it once again for a medium consisting
of a mixture of different particle species which interact with each other
both elastically and inelastically:
\begin{equation}
\label{eq1}
\tau_\mathrm{exp}(x) \equiv \frac{1}{\partial\cdot u(x)} = 
\xi\, \tau^{(i)}_\mathrm{scatt}(x) \equiv \xi\, \frac{1}{\sum_j
\langle\sigma_{ij}v_{ij}\rangle\rho_j(x)},
\end{equation}
where $\xi$ is an (unknown) parameter of order 1. Here we exhibit not 
only the already mentioned dependence of the criterium on the particle 
species $i$, but also the fact that it is a {\em local} criterium, 
depending on the local fluid expansion rate $\partial\cdot u(x)$
and on the local scattering rate $\sum_j\langle\sigma_{ij}v_{ij}\rangle
\rho_j(x)$ (where $\sigma_{ij}$ is the scattering cross section between 
particle species $i$ and $j$ and $v_{ij}$ their relative velocity in the 
pair center of mass frame). Thus, different parts of the expanding 
fireball will, in general, decouple at different times, depending on 
the flow velocity and density profiles $u(x)$ and $\rho_j(x)$. The 
ensemble of points $(\bm{x},\tau_f(\bm{x}))$ satisfying Eq.~(\ref{eq1}) 
defines the \textbf{\em freeze-out hypersurface}. It is a 3-dimensional 
surface imbedded in 4-dimensional space-time. 

Since chemical (particle number changing) reactions involve different 
types of cross sections than thermal (momentum changing) reactions,
equality of the two sides of the equation will be satisfied on different
freeze-out surfaces: Chemical reactions exploit only a small fraction
$\sigma_{ij}^\mathrm{inel}$ of the total transport cross section 
$\sigma_{ij}^\mathrm{tot}$ corresponding to all possible momentum 
changing processes. Since  $\sigma_{ij}^\mathrm{inel} < 
\sigma_{ij}^\mathrm{tot}$, the mean free time $\tau_\mathrm{scatt}$ 
between two inelastic scattering processes is longer than that between
two arbitrary momentum changing processes. Consequently, chemical processes
decouple before the elastic scattering processes, and the hadron abundances 
freeze out earlier than the momentum spectra \cite{Heinz:1993cj}: 
\begin{equation}
\label{eq2}
T^{(i)}_\mathrm{chem} > T^{(i)}_\mathrm{therm}.
\end{equation}
Since the scattering rate is particle specific, different hadrons should 
still freeze out at different temperatures, both chemically and thermally.
To implement differential {\em thermal} freeze-out into a hydrodynamic 
model for the fireball expansion is, however, difficult since it would
require the introduction of sophisticated loss terms describing the
decoupling of the particles from the fluid. Fortunately, usually there is
one species that dominates the scattering cross section (nucleons
at low collision energies, pions at high collision energies) whose 
freeze-out triggers all others.

\vspace*{-3mm}
%%%%%%%%%%%%%%%%%%%%%%%%%%%%%%%%%%%%%%%%%%%%%%%%%%%%%%%%%%%%%%%%%%%%%%%
\section{Chemical and thermal freeze-out and the QCD phase transition}
\label{sec3}
\subsection{The experimental situation anno 2000}
\label{sec3.1}
\begin{figure}[t]
\begin{center}
\includegraphics*[width=0.7\columnwidth,height=9.5cm,clip=]%
                 {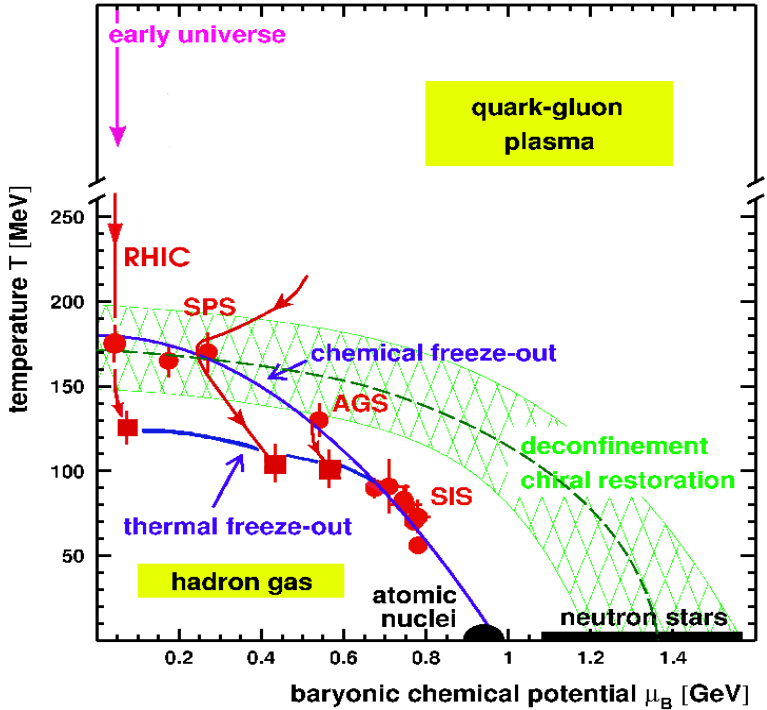}
\end{center}
\caption{Chemical and thermal freeze-out points extracted from heavy-ion
         collisions at the GSI SIS, BNL AGS, CERN SPS and RHIC. The shaded
         area indicates the likely location of the quark-hadron phase 
         transition as extracted from lattice QCD and theoretical
         models. An updated version of the chemical freeze-out points
         can be found in \cite{Cley}.}
\label{F2}
\end{figure}

%%%%%%%%%%%%%%%%%%%%%%%%%%%%%%%%%%%%%%%%%%%%%%%%%%%%%%%%%%%%%%%%%%%%%%%
%
In 2000, after 15 years of fixed-targed collision experiments with
ultra-relativistic heavy-ion beams at the Brookhaven AGS and CERN SPS,
the Relativistic Heavy Ion Collider RHIC began colliding countercirculating
beams of Au ions at BNL. Following the theoretical concepts developed 
during this period and outlined above, a large body of data had been 
collected on chemical freeze-out of hadron yields and thermal freeze-out 
of their momentum spectra at different collision energies. The hadrons 
emitted in relativistic heavy-ion collisions show 
thermal characteristics both in their abundances and in the shapes of 
their transverse momentum spectra. The status of these analyses shortly 
after RHIC turned on is depicted in Fig.~\ref{F2}.

A short summary of the observations made by the various groups that 
contributed to this compilation is as follows: (i) Above AGS energies,
the extracted chemical and thermal freeze-out temperatures clearly
differ from each other, with chemical decoupling occurring at higher
temperature \cite{Cleymans:1998fq,BRS03,STAR_Tchem,Dobler:1999ju,%
Tomasik:1999cq,JBH03,STAR_Tdec}, just as predicted by Jozs\'o's 
dynamical freeze-out criterium. (ii) For $\sqrt{s}\gtrsim 10\,A$\,GeV, 
one observes the same ``universal'' chemical freeze-out temperature 
$T_\mathrm{chem}$ in $e^+e^-$, $pp$, $p\bar p$, and $AA$ collisions 
\cite{Becattini:1997rv,Bec} (the only difference between small and 
large collision systems being the level of strangeness saturation). 
(iii) For $\sqrt{s}\gtrsim 10\,A$\,GeV, $T_\mathrm{chem}$ agrees with 
the critical temperature $T_c$ for the color-confining quark-hadron 
phase transition predicted by lattice QCD for (approximately) 
baryon-free hot hadronic matter \cite{Karsch:2001cy,Aoki:2006br}. 
(Recent developments along this front will be discussed further below.) 
(iv) Hadronic cascades (RQMD, UrQMD, \dots) show that hadronic 
rescattering after the QGP hadronization alters the momentum 
distributions and resonance populations of the hadrons (thereby 
cooling the system while keeping it -- at least for a while -- close 
to local thermal equilibrium \cite{Bravina}), but not the stable 
hadron yields \cite{Bass:2000ib,H99}, again in agreement with the 
above considerations based on Jozs\'o's dynamical freeze-out criterium.
[Hadronic rescattering leads to the loss of a fraction of the 
baryon-antibaryon pairs, but this can be at least partially traced back 
to the absence of multi-hadron collision channels so that detailed balance 
is violated in baryon-antibaryon annihilation \cite{RS01}.]
%
%%%%%%%%%%%%%%%%%%%%%% Fig. 2 %%%%%%%%%%%%%%%%%%%%%%%%%%%%%%%%%%%%%%%%%%%
%%%%%%%%%%%%%%%%%%%%%%%%%%%%%%%%%%%%%%%%%%%%%%%%%%%%%%%%%%%%%%%%%%%%%%%%%
%  

\vspace*{-2mm}
%%%%%%%%%%%%%%%%%%%%%%%%%%%%%%%%%%%%%%%%%%%%%%%%%%%%%%%%%%%%%%%%%%%%%%%%
\subsection{The controversy: Kinetic freeze-out of chemical reactions
or statistical hadronization?}\label{sec3.2}
%%%%%%%%%%%%%%%%%%%%%%%%%%%%%%%%%%%%%%%%%%%%%%%%%%%%%%%%%%%%%%%%%%%%%%%%

These empirical facts have split the heavy-ion theory community into
two camps which offer different interpretations of the observations.
The philosophy of {\bf Camp I} is laid out in Refs.~\cite{Becattini:1997rv,%
H99,H98,S99} and holds that hadron production is a statistical process 
associated with a phase transition, proceeding through very many 
different possible microscopic channels constrained only by energy, 
baryon number and (both net and total) strangeness conservation, 
thereby leading to a maximum entropy (i.e. statistically most 
probable) configuration described by a thermal distribution of 
hadron yields. In this interpretation the extracted ``thermodynamic'' 
parameters $T_\mathrm{chem}$, $\mu_B$ and $\gamma_s$ play the role of 
Lagrange multipliers to ensure these conservation law constraints
while maximizing the entropy \cite{Slotta:1995bh}. The value of 
$T_\mathrm{chem}$ is {\em not} established by inelastic reactions 
among hadrons proceeding until chemical equilibrium is reached  -- 
rather, the hadrons are directly ``born'' into a maximum entropy 
state of apparent chemical equilibrium \cite{S99}, with the parameter 
$T_\mathrm{chem}$ defining the critical energy density $e_c$
at which the hadronization process happens (\cite{Becattini:1997rv}). 
$T_\mathrm{chem}$ is thus conceptually different from the thermal (or
kinetic) decoupling temperature $T_\mathrm{therm}\equiv T_\mathrm{kin}$ 
which {\em is} the result of quasi-elastic rescatterings among the 
hadrons (which also contribute to their collective flow).  

{\bf Camp II} includes the followers of Refs.~\cite{RS01,BSW04,GrSQM04}
who hold that chemical freeze-out is a kinetic process within the hadronic
phase, conceptually equivalent with kinetic freeze-out, the only difference
being the quantitative values of the corresponding freeze-out temperatures
which reflect the fact that the inelastic cross sections driving chemical 
equilibration constitute only a small fraction of the total scattering
cross section contributing to momentum exchange. The hadrons are not born
into chemical equilibrium, but driven into such a state kinetically by
inelastic multi-hadron processes (which, according to 
Refs.~\cite{RS01,BSW04,GrSQM04}, become crucial near $T_c$ due to 
high hadron densities) and frozen out by global expansion. Accordingly,
$T_\mathrm{chem}$ is the ``real'' temperature describing the latest 
point at which forward and backward chemical reactions balance each 
other. (In contrast, for Camp I, there are no ``backward'' reactions 
involving hadrons in both initial and final states.)

%%%%%%%%%%%%%%%%%%%%%%%%%%%%%%%%%%%%%%%%%%%%%%%%%%%%%%%%%%%%%%%%%%%%%%%%
\subsection{How to resolve the controversy: RHIC precision data}
\label{sec3.3}
%%%%%%%%%%%%%%%%%%%%%%%%%%%%%%%%%%%%%%%%%%%%%%%%%%%%%%%%%%%%%%%%%%%%%%%%
%
Is this more than a philosophical difference of opinions? We think so --
this controversy can be resolved unambiguously \cite{Heinz:2006ur}. To 
better explain our argument let us first cast a more detailed look at the 
recent precision data collected at RHIC.
%
%%%%%%%%%%%%%%%%%%%%%% Fig. 3 %%%%%%%%%%%%%%%%%%%%%%%%%%%%%%%%%%%%%%%%%%%
\begin{figure}[htb]
\includegraphics*[bb=0 -6 543 375,width=0.4\columnwidth,height=4.95cm,clip=]%
                 {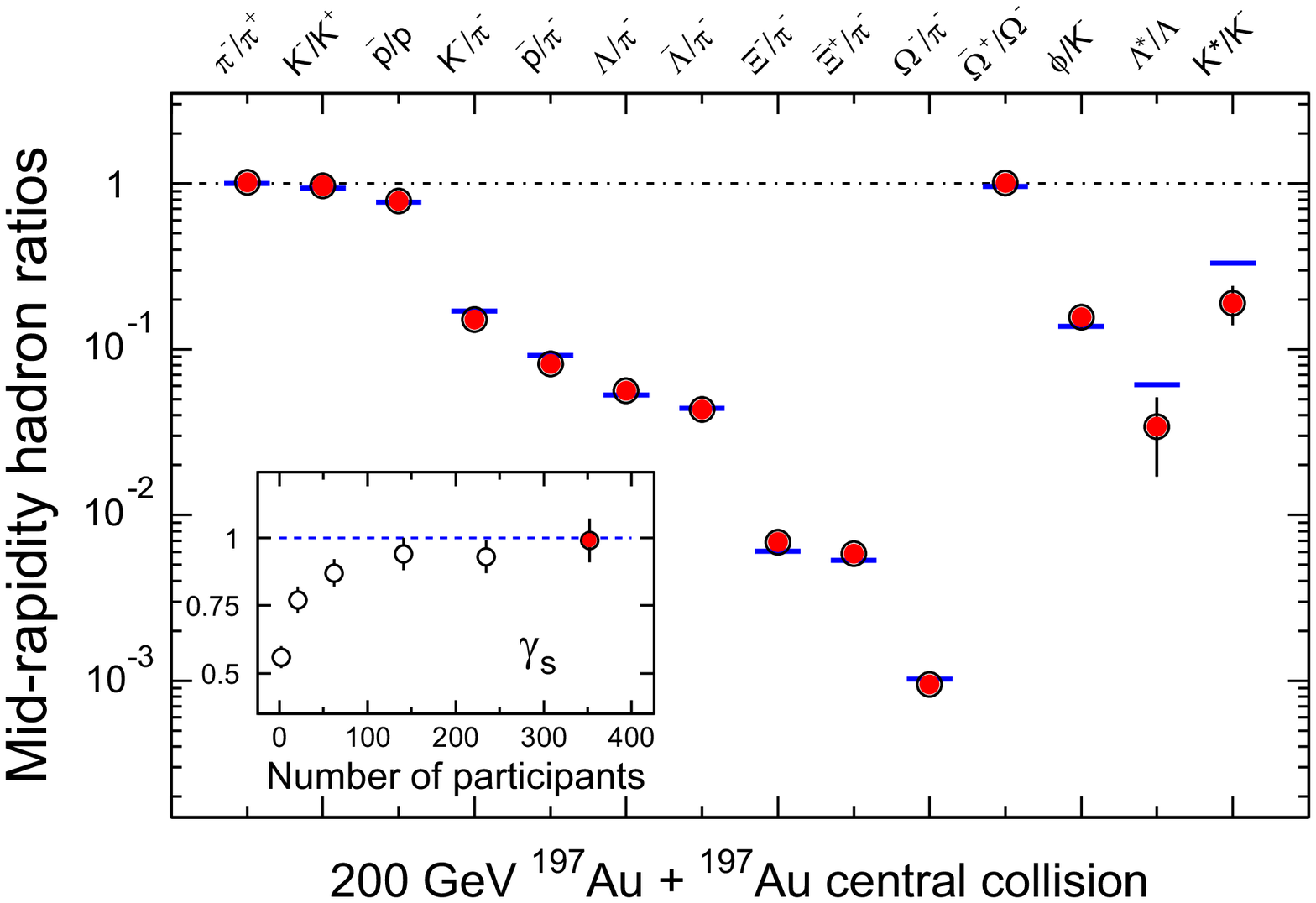}
\includegraphics*[bb=0 10 567 400,width=0.6\columnwidth,height=5cm]%
                 {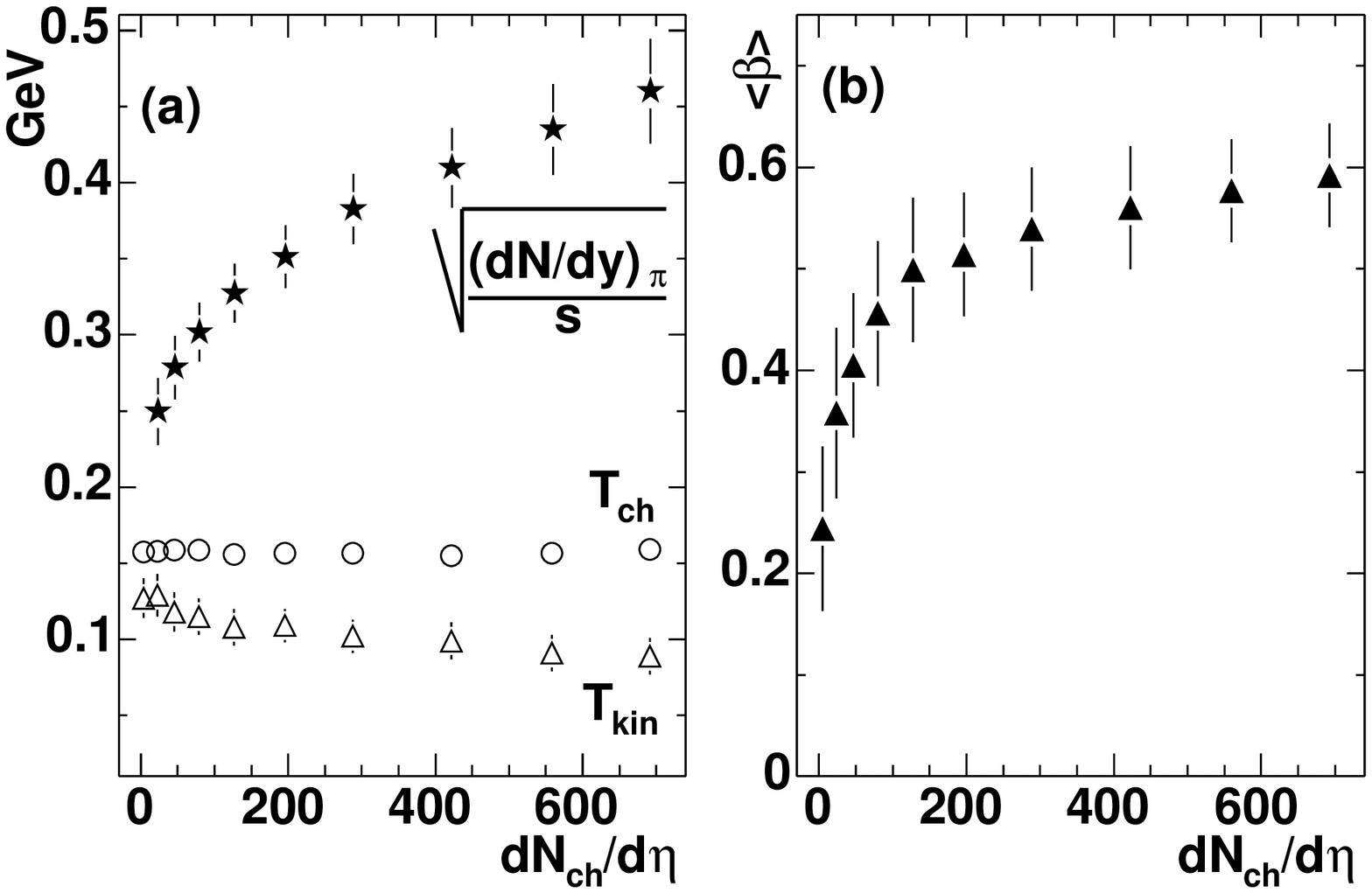}
\caption{{\sl Left:} Abundance ratios of stable hadrons from central
         $200\,A$\,GeV Au+Au collisions at RHIC \cite{STAR_Tchem}. The inset
         shows the centrality dependence of the strangeness saturation 
         factor $\gamma_s$. 
         {\sl Right:} Centrality dependence (with centrality measured by 
         charged hadron rapidity density $dN_\mathrm{ch}/d\eta$) of (a) 
         the thermal freeze-out temperature 
         $T_\mathrm{kin}{\equiv}T_\mathrm{therm}$ (open triangles), the
         chemical freeze-out temperature $T_\mathrm{chem}$ (open circles),
         and the square root of the transverse areal density of pions
         $(dN_\pi/d\eta)/S$ (solid stars), and (b) the average transverse 
         flow velocity $\langle\beta\rangle{\equiv}\langle v_\perp\rangle$
         (solid triangles), for the same collision system \cite{STAR_Tdec}.}
\label{F3}
\end{figure}
%%%%%%%%%%%%%%%%%%%%%%%%%%%%%%%%%%%%%%%%%%%%%%%%%%%%%%%%%%%%%%%%%%%%%%%%%
%  
Figure~\ref{F3} shows the results from thermal model fits to hadron yield
ratios and transverse momentum spectra from Au+Au collisions at 
$\sqrt{s}=200\,A$\,GeV. The final hadron abundances from central 
collisions can be described by a hadron resonance gas in a state of 
approximate chemical equilibrium at $T_\mathrm{chem}=163\pm4$\,MeV, 
$\mu_B=24\pm4$\,MeV, and a strangeness saturation factor 
$\gamma_s=0.99\pm0.07$ \cite{STAR_Tchem}. The quality of the statistical 
model fit is impressive. The STAR collaboration also studied the 
dependence of the fit parameters on the collision centrality and found 
that neither the temperature $T_\mathrm{chem}$ nor the baryon chemical 
potential $\mu_B$ depend appreciably on the impact parameter 
\cite{STAR_Tdec} \cite{fn2}; only the strangeness suppression factor 
exhibits centrality dependence, beginning at impact parameters $>8-9$\,fm, 
and drops to values around 0.55 in the most peripheral Au+Au collisions 
\cite{STAR_Tchem}. The centrality independence of $T_\mathrm{chem}$ 
(open circles in the middle panel of Fig.~\ref{F3}) is in stark contrast 
to the behavior observed in the same experiment for the kinetic (thermal)
decoupling temperature $T_\mathrm{kin}{\equiv}T_\mathrm{therm}$, which is 
extracted together with a value for the average radial flow velocity 
$\langle\beta\rangle$ of the fireball at thermal freeze-out from the 
shape of the transverse momentum spectra of identified pions, kaons and 
(anti-)protons \cite{STAR_Tdec}: Thie right two panels in Fig.~\ref{F3} 
show that $T_\mathrm{kin}$ increases significantly with increasing impact 
parameter, from $T_\mathrm{kin}=89\pm12$\,MeV in the most central to   
$T_\mathrm{kin}=127\pm13$\,MeV in the most peripheral collisions, while
at the same time the average radial flow decreases from 
$\langle\beta\rangle=0.59\pm0.05$ in the most central to
$\langle\beta\rangle=0.24\pm0.08$ in the most peripheral Au+Au collisions.
This last observation demonstrates a strong centrality dependence of the 
fireball expansion dynamics.

Returning to the controversy between Camps I and II described in the
preceding subsection, we note that Camp II has to cope with an intrinsic 
tension between two observations: The high quality of the thermal model 
fit to the observed hadron yields at RHIC requires sufficient time for 
inelastic reactions to establish a good chemical equilibrium, whereas 
the proximity of the fitted chemical freeze-out temperature 
$T_\mathrm{chem}$ to the critical temperature $T_c$ of the quark-hadron 
phase transition from lattice QCD, together with the rapid cooling of 
the fireball by collective expansion, don't provide much of a time window 
for these processes to play out. In essence, to make the kinetic chemical 
equilibration scenario work one needs {\em very} large scattering rates 
right near $T_c$ which then drop to negligible values just below $T_c$. 
[This would be easier to understand if there were a larger gap between 
$T_c$ and $T_\mathrm{chem}$, as suggested by the recent upward revision 
of $T_c$ from Lattice QCD advocated in \cite{Karsch_Tc}, but this problem
appears serious if the lower $T_c(\chi_{\bar\psi\psi})$ from 
Ref.~\cite{Aoki:2006br} turns out to be correct.] 

In addition, we point out a second conceptual problem with the 
Camp II interpretation: If freeze-out is a kinetic process, it
is controlled by the competition between local scattering (moving
the system towards equilibrium) and global expansion (driving the 
system out of equilibrium). The resulting freeze-out temperature
is therefore sensitive to the fireball expansion rate which (as the
right panel in Fig.~\ref{F3} shows) depends on collision centrality.
Thus the extracted kinetic decoupling temperature should also depend
on centrality. While such a centrality dependence is empirically 
observed for the kinetic decoupling temperature $T_\mathrm{kin}$, the 
chemical freeze-out temperature does not appear to vary with collision 
centrality (middle panel in Fig.~\ref{F3}). Hence it cannot be the result 
of a kinetic decoupling process from inelastic hadronic scattering.

According to Jozs\'o Zim\'anyi's dynamical freeze-out criterium, 
dependence of the freeze-out temperature on the collective expansion 
rate, and through this rate on the collision centrality, is a 
{\em tell-tale signature for a kinetic decoupling process.} In the rest 
of this paper we show that the observed centrality dependences of the 
average radial flow velocity and thermal freeze-out temperature are 
consistent with hydrodynamic behaviour of the fireball medium followed by 
kinetic decoupling of the hadrons from microscopic scattering processes, 
driven by the collective expansion. We will then show that a centrality
independent freeze-out temperature is inconsistent with a kinetic 
decoupling process unless the chemical scattering rates have an 
extremely (i.e. almost infinitely) strong temperature dependence. We 
therefore interpret the observed centrality independence of 
$T_\mathrm{chem}$ as evidence that chemical decoupling of the hadron 
abundances is driven by a {\em phase transition} during which the chemical 
reaction rates decrease precipitously, leaving the system in a chemically 
frozen-out state at the end of the transition. Only in this way 
is it possible to obtain a universal chemical freeze-out temperature 
that is {\em insensitive} to the (centrality dependent) collective 
dynamics, and only depends on the thermodynamic parameters of the phase 
transition. Obviously, the chemical processes happening during the 
hadronization process itself involve colored degrees of freedom and 
can thus not be efficiently described in hadronic language. We also 
address the centrality dependence of the strangeness saturation factor 
and comment on how our picture also reproduces chemical abundance 
data measured in $pp$ and $e^+e^-$ collisions.  

%%%%%%%%%%%%%%%%%%%%%%%%%%%%%%%%%%%%%%%%%%%%%%%%%%%%%%%%%%%%%%%%%%%%%%%%
\section{Kinetic freeze-out from a hydrodynamically expanding system}
\label{sec4}
%%%%%%%%%%%%%%%%%%%%%%%%%%%%%%%%%%%%%%%%%%%%%%%%%%%%%%%%%%%%%%%%%%%%%%%%

In this section we show that the hydrodynamic model can quantitatively 
reproduce the observed centrality dependence of the kinetic decoupling
temperature extracted from hadron momentum spectra at RHIC. We then show
that an analogous centrality dependence of the chemical freeze-out 
temperature cannot be avoided if the hadron yields are similarly controlled 
by kinetic freeze-out from inelastic hadronic rescattering. 

%%%%%%%%%%%%%%%%%%%%%%%%%%%%%%%%%%%%%%%%%%%%%%%%%%%%%%%%%%%%%%%%%%%%%%%%%
\subsection{Kinetic thermal freeze-out from hydrodynamics}\label{sec4.1}
%%%%%%%%%%%%%%%%%%%%%%%%%%%%%%%%%%%%%%%%%%%%%%%%%%%%%%%%%%%%%%%%%%%%%%%%%
%
We use our (2+1)-dimensional longitudinally boost-invariant hydrodynamic
code AZHYDRO \cite{AZHYDRO} with standard initial conditions \cite{QGP3} 
to generate the flow pattern for $200\,A$\,GeV Au+Au collisions. This code
has been previously shown to successfully reproduce the measured single 
particle hadron $p_T$-spectra and their elliptic flow (for details see
\cite{H_SQM04}). Here, however, we modify the freeze-out criterium for 
thermal decoupling to account for its kinetic nature: Instead of requiring
freeze-out on a surface of constant energy density $e_\mathrm{dec} =
0.075$\,GeV/fm$^3$ (corresponding to a fixed temperature 
$T_\mathrm{kin}=100$\,MeV \cite{QGP3}), we define the kinetic freeze-out 
surface as the set of points satisfying Eq.~(\ref{eq1})
\cite{BGZ78,Schnekki,Hung}. In a first attempt, the proportionality 
constant is set to $\xi=0.35$, yielding an average temperature along the 
freeze-out surface for central Au+Au collisions of 
$\langle T_\mathrm{kin}\rangle\simeq 115$\,MeV. Having fixed $\xi$ in 
central collisions, Eq.~(\ref{eq1}) is taken to define the freeze-out 
surface also at other impact parameters. Inside the freeze-out surface 
the scattering rate exceeds $\xi$ times the expansion rate, and the matter 
is thermalized; outside the surface the expansion rate exceeds 
$\xi^{-1}$ times the scattering rate -- there the hadrons are assumed to 
be decoupled from the fluid, streaming freely into the detector.

The expansion rate $\partial\cdot u=\gamma_\perp\left(\frac{1}{\tau}
+\bm{\nabla}_\perp\cdot\bm{v}_\perp\right) + \left(\partial_\tau + 
\bm{v}_\perp{\cdot}\bm{\nabla}_\perp\right) \gamma_\perp$ is computed
from the hydrodynamic output for the transverse flow velocity 
$\bm{v}_\perp(x)$ and $\gamma_\perp=(1{-}v_\perp^2)^{-1/2}$. Since at 
RHIC energies hadron production is dominated by pions, we assume for 
simplicity that all hadrons decouple when pions freeze out. The pion 
scattering rate is taken from the numerical results presented in 
Ref.~\cite{Hung} which we parametrize as
\begin{equation}
\label{rate}
  \frac{1}{\tau^\pi_\mathrm{scatt}} = \left(59.5\,\mathrm{fm}^{-1}\right)
  \left(\frac{T}{1\,\mathrm{GeV}}\right)^{3.45}.
\end{equation}
This defines the rate for momentum changing collisions, to be used for 
the calculation of thermal freeze-out, and will need to be modified 
when discussing chemical freeze-out below.

%
%%%%%%%%%%%%%%%%%%%%%%%%%%%%%%%%%%%%% Fig. 4 %%%%%%%%%%%%%%%%%%%%%%%%%%%%%%%
\begin{figure}[htb]
%\centerline{\epsfig{file=dndm2_0.ps,bb=51  208  554  575,width=7.9cm}}
\begin{center}
\includegraphics*[bb=50 50 554 445,width=8.5cm,height=5.5cm,clip=]{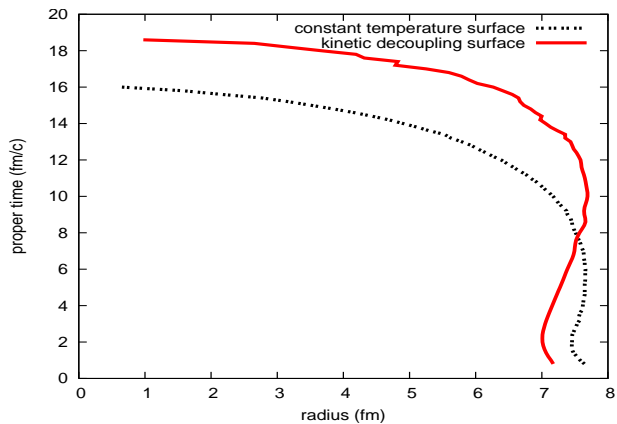}
\end{center}
\caption{Kinetic (thermal) freeze-out surface $\tau_\mathrm{kin}(r)$ for 
central ($b=0$) $200\,A$\,GeV Au+Au collisions, computed from Eq.~(\ref{eq1}) 
with $\xi=0.35$ (red solid line) and for a constant freeze-out temperature
$T_\mathrm{kin}=115$\,MeV (dotted black line). Both surfaces have the
same average temperature of $\langle T\rangle =115$\,MeV, using the 
energy density as weight function.
\label{F4}}
\end{figure}
%%%%%%%%%%%%%%%%%%%%%%%%%%%%%%%%%%%%%%%%%%%%%%%%%%%%%%%%%%%%%%%%%%%%%%%%%%%%
%

In Fig.~\ref{F4} we plot the kinetic freeze-out surface for central Au+Au
collisions computed from Eq.~(\ref{eq1}) with $\xi=0.35$ (solid red line)
and from the condition $T_\mathrm{kin}=115$\,MeV (dotted black line). Both
have the same average kinetic freeze-out temperature, but for the kinetic 
freeze-out criterium (\ref{eq1}) the middle of the fireball freezes out
a bit later at lower temperature and larger flow whereas the edge decouples
earlier at higher temperature and with less flow than the contant-$T$ 
surface. This is caused by the larger expansion \emph{rate} near the
edge of the fireball.

%
%%%%%%%%%%%%%%%%%%%%%%%%%%%%%%%%%%%%% Fig. 5 %%%%%%%%%%%%%%%%%%%%%%%%%%%%%%%
\begin{figure}[thb]
%\centerline{\epsfig{file=dndm2_0.ps,bb=51  208  554  575,width=7.9cm}}
\includegraphics*[width=7.6cm,height=5.5cm]{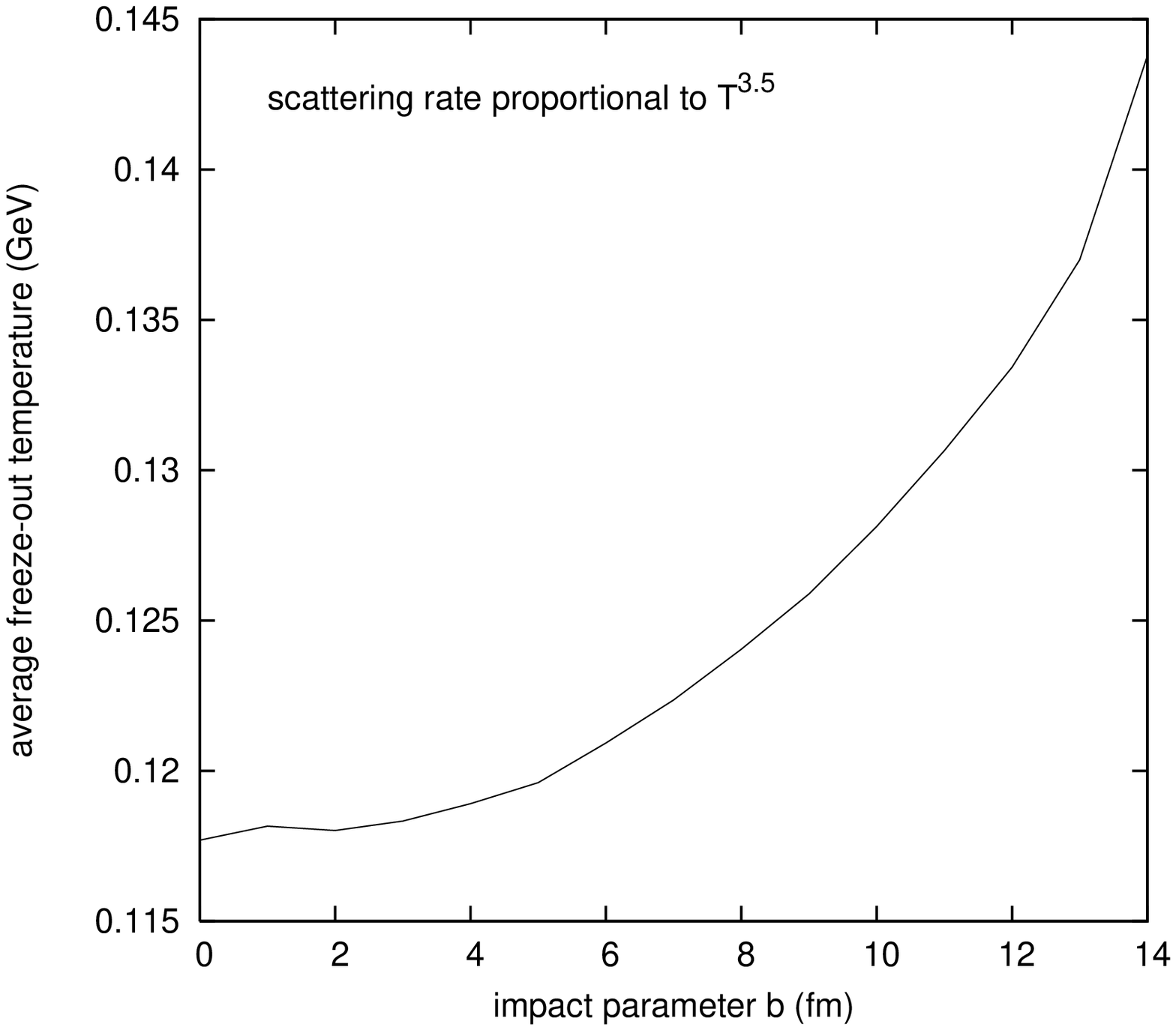}
\includegraphics*[width=7.7cm,height=5.5cm]{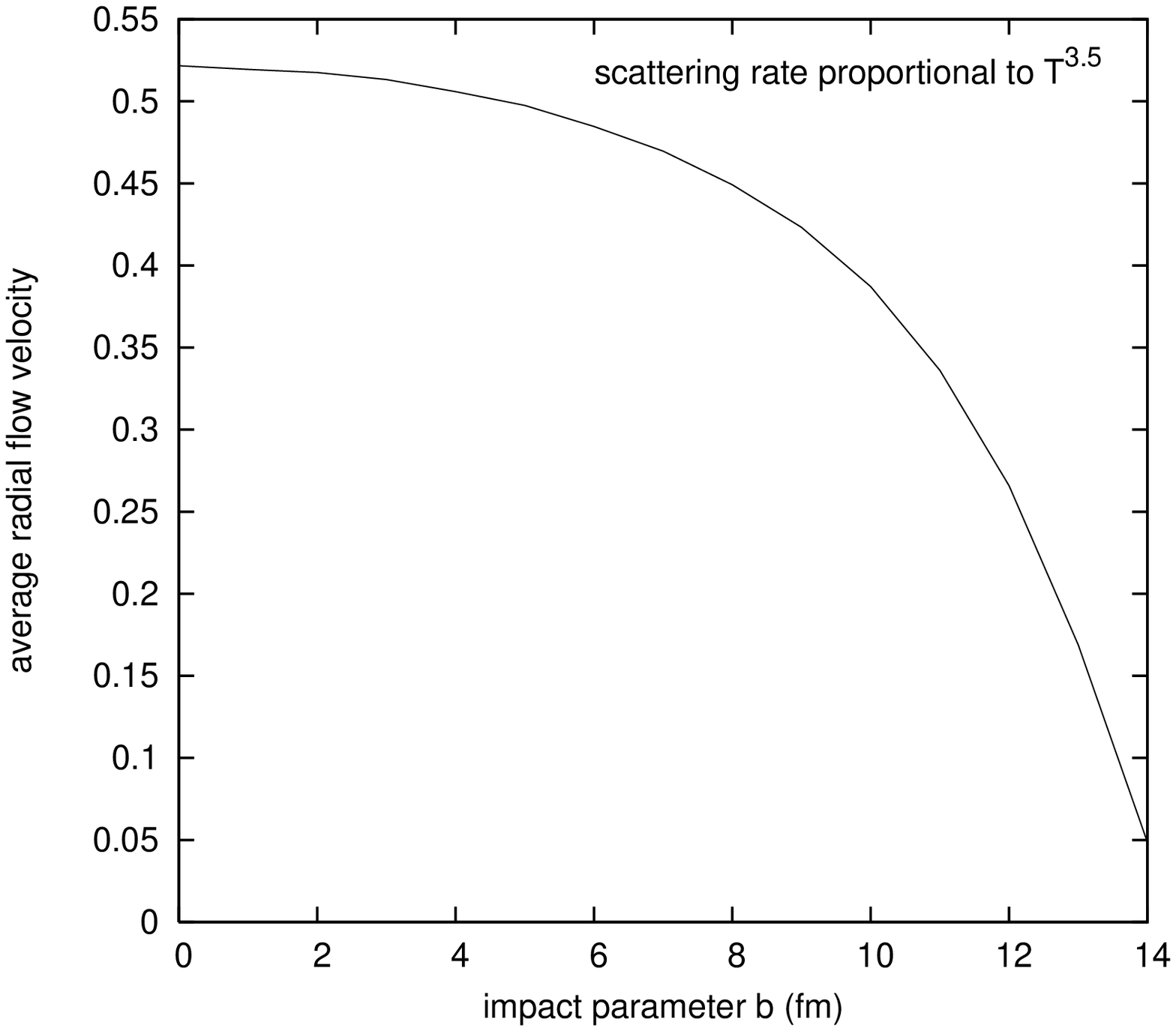}
\caption{Impact parameter dependence of the average kinetic decoupling
temperature $\langle T_\mathrm{kin}\rangle$ (left) and average radial 
flow velocity $\langle v_\perp\rangle$ along the freeze-out surface
computed from Eq.~(\protect\ref{eq1}) with $\xi=0.35$ and a scattering 
rate $\sim T^{3.45}$ as in Eq.~(\protect\ref{rate}), for $200\,A$\,GeV
Au+Au collisions.\label{F5}}
\end{figure}
%%%%%%%%%%%%%%%%%%%%%%%%%%%%%%%%%%%%%%%%%%%%%%%%%%%%%%%%%%%%%%%%%%%%%%%%%%%%
%
Figure~\ref{F5} shows the impact parameter dependence of the average kinetic
decoupling temperature and the associated average radial flow calculated
from the kinetic freeze-out criterium (\ref{eq1}) with $\xi=0.35$. Central 
collisions are seen to decouple at relatively low temperatures with
large average flow whereas peripheral collisions freeze out earlier
when the fireballs are still hotter and less flow has developed. [Note 
that the average flow velocity is {\em smaller} in peripheral collisions, 
but the expansion {\em rate} (i.e. the flow velocity {\em gradient}) is 
{\em larger}\,!] This is in good qualitative agreement with the STAR data
\cite{STAR_Tdec}, although their freeze-out temperatures are generally a 
bit lower, with slightly larger average radial flow velocities than seen 
in Fig.~\ref{F5}. 

\begin{figure}[htb]
%\centerline{\epsfig{file=dndm2_0.ps,bb=51  208  554  575,width=7.9cm}}
\begin{center}
\includegraphics*[width=9cm,height=6cm]{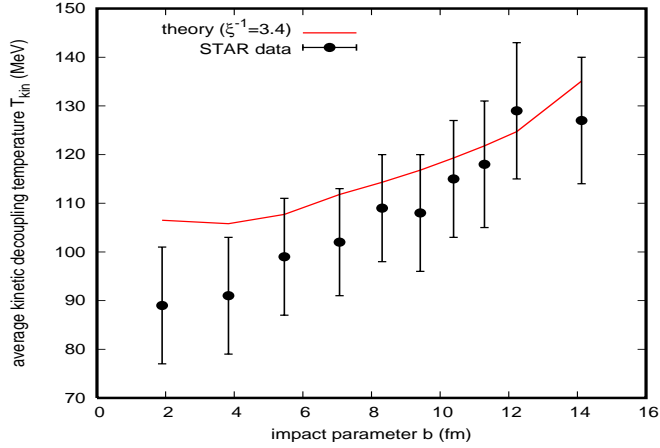}
\end{center}
\caption{Impact parameter dependence of the average kinetic decoupling
temperature $\langle T_\mathrm{kin}\rangle$ computed from hydrodynamics with
kinetic freeze-out criterium (\protect\ref{eq1}) using $\xi=0.295$, 
compared with STAR data \protect\cite{STAR_Tdec} for $200\,A$\,GeV
Au+Au collisions.
\label{F6}}
\end{figure}

We adjust for this by fine-tuning the phenomenological parameter $\xi$
in Eq.~(\ref{eq1}) to $\xi=0.295$ ($\xi^{-1}=3.4$). The corresponding
freeze-out temperatures are shown as a function of impact parameter $b$
in Figure~\ref{F6}, together with the STAR data.
Now the agreement is also quantitatively acceptable. We conclude that the 
measured centrality dependence of $T_\mathrm{kin}$ can be completely 
understood in terms of a hydrodynamic model for the fireball expansion, 
coupled to a kinetic freeze-out criterium with realistic temperature 
dependence of the microscopic scattering rate.  
%
%%%%%%%%%%%%%%%%%%%%%%%%%%%%%%%%%%%%% Fig. 6 %%%%%%%%%%%%%%%%%%%%%%%%%%%%%%%
%%%%%%%%%%%%%%%%%%%%%%%%%%%%%%%%%%%%%%%%%%%%%%%%%%%%%%%%%%%%%%%%%%%%%%%%%%%%
%

\vspace*{-3mm}
%%%%%%%%%%%%%%%%%%%%%%%%%%%%%%%%%%%%%%%%%%%%%%%%%%%%%%%%%%%%%%%%%%%%%%%%%
\subsection{Kinetic chemical freeze-out from hydrodynamics?}\label{sec4.2}
%%%%%%%%%%%%%%%%%%%%%%%%%%%%%%%%%%%%%%%%%%%%%%%%%%%%%%%%%%%%%%%%%%%%%%%%%
%
Let us now see whether we can similarly understand chemical freeze-out
as a kinetic decoupling process from {\em inelastic} hadronic scattering. A
few typical processes relevant for chemical equilibration are
\begin{eqnarray}
\label{inel}
&&\pi + \pi \longleftrightarrow K + \bar K, \qquad
  \pi + N \longleftrightarrow K + Y,, \qquad
  \pi + Y \longleftrightarrow \bar K + N,  
\nonumber\\
&&\Omega + \pi \longleftrightarrow \Xi + \bar K,\qquad
  K + \bar K \longleftrightarrow \phi + \pi,\qquad
  \Omega + \bar K \longleftrightarrow \Xi + \pi,
\\
&&\Omega + \bar N \longleftrightarrow 2 \pi + 3\bar K,\qquad
  N + \bar N \longleftrightarrow 5\pi,\qquad
  N + 3\bar K \longleftrightarrow \Omega + 3\pi.
%  \Xi + \bar Y \longleftrightarrow 4\pi + \bar K.
\nonumber
\end{eqnarray}
The last line shows so-called multi-hadron collision channels which, in at 
least one direction, require collisions between more than two hadrons.
Rates for processes involving $n_\mathrm{in}$ incoming hadrons are
proportional to the product of their densities $\sim \Pi_{i=1}^{n_\mathrm{in}}
n_i(T)$ where each factor $n_i(T)$ grows with $T$ at least as $T^3$ (even 
much more rapidly for hadrons with masses $>T$). At low temperatures, 
multi-hadron collision processes as well as collisions between very massive 
hadrons are therefore strongly suppressed. Consequently, particle yields for 
hadrons requiring collisions of many abundantly available particles for their 
production or destruction (such as $\bar p, \Omega, \dots$) thus tend to 
freeze out at higher $T$ than particle yields for hadrons whose abundances 
can be efficiently changed by two-body reactions ($\pi, K, \phi, \dots$).

In an expanding, cooling system, simultaneous freeze-out of all hadron 
yields at a {\em common} temperature therefore requires a {\em conspiracy 
of rates} with widely differring $T$-dependences. Indeed, thermal model fits 
to hadron abundances with a single common temperature are usually not
perfect \cite{SHSX98}, and individual fits to subsets of yields measured
in lower-energy collisions at the SPS and AGS tend to lead to a significant 
spread of chemical freeze-out temperatures \cite{DPZ06}. So far, only at 
RHIC does the single-temperature chemical equilibrium fit give an almost 
perfect description of the data \cite{BRS03,DPZ06}.

One way to achieve the conspiracy of different chemical equilibration 
rates that is required for a good fit with a single freeze-out temperature 
is to postulate that at chemical freeze-out {\em all} chemical reactions
are completely dominated by multi-hadron collisions and that at any
temperature below $T_\mathrm{chem}$ the medium is so rarefied and so 
rapidly expanding that even the simplest two-body reactions among the
most abundantly produced hadrons (such as those listed in the first line
of Eq.~(\ref{inel})) have essentially stopped. As long as collision 
channels with widely different temperature dependences compete with
each other, chemical freeze-out of all hadron species at a single 
temperature appears to be impossible.

Even more importantly, even if it were possible at one fixed impact 
parameter to arrange for common freeze-out of all hadron species in 
spite of a competition of scattering rates with different temperature
dependences, such a conspiracy would be impossible to maintain, {\em 
with the same value for the freeze-out temperature $T_\mathrm{chem}$},
%
%%%%%%%%%%%%%%%%%%%%%%%%%%%%%%%%%%%%% Fig. 7 %%%%%%%%%%%%%%%%%%%%%%%%%%%%%%%
\begin{figure}[ht]
%\centerline{\epsfig{file=dndm2_0.ps,bb=51  208  554  575,width=7.9cm}}
\begin{center}
\includegraphics*[width=7cm,height=10cm,angle=270]{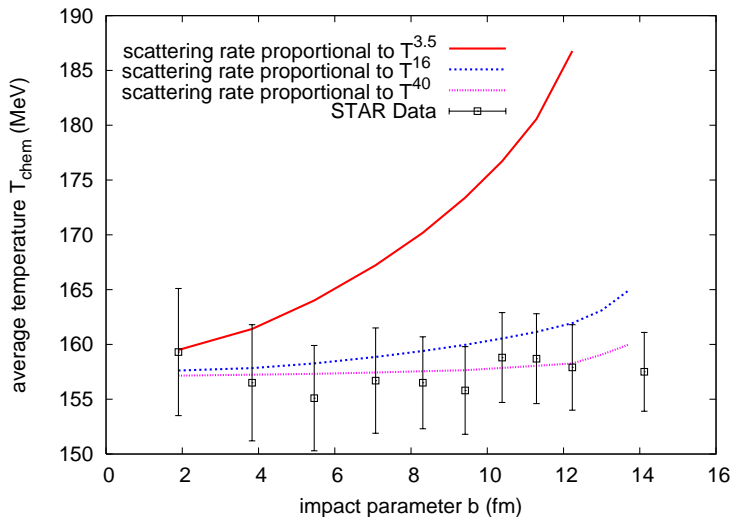}
\end{center}
\caption{Impact parameter dependence of the average chemical decoupling
temperature $\langle T_\mathrm{chem}\rangle$ computed from hydrodynamics 
with kinetic freeze-out criterium (\protect\ref{eq1}) using $\xi=0.95$
and reaction rates with different temperature dependences as listed, 
compared with STAR data \protect\cite{STAR_Tdec} for $200\,A$\,GeV
Au+Au collisions.
\label{F7}}
\end{figure}
%%%%%%%%%%%%%%%%%%%%%%%%%%%%%%%%%%%%%%%%%%%%%%%%%%%%%%%%%%%%%%%%%%%%%%%%%%%%
%
over the entire impact parameter range. Figure~\ref{F7} shows the 
centrality dependence of the average chemical freeze-out temperatures 
along hydrodynamic decoupling surfaces computed with the kinetic 
freeze-out criterium (\ref{eq1}), using $\xi=0.95$ to adjust the 
value of $\langle T_\mathrm{chem}\rangle$ in central Au+Au collisions 
to the STAR data \cite{STAR_Tdec} and exploring different possible 
temperature dependences of the dominant inelastic scattering rate. One 
sees that approximate impact parameter independence of 
$\langle T_\mathrm{chem}\rangle$ can only be achieved if \emph{all} 
inelastic scattering rates grow with $T$ as $T^n$ with a power 
$n\gtrsim20$!

Basically, Fig.~\ref{F7} tells us that the observed centrality
independence of $T_\mathrm{chem}$ requires chemical freeze-out to
happen in a region of parameter space where {\em all} chemical 
reaction rates exhibit extremely steep temperature dependence, dropping
like a stone as the system cools through the decoupling temperature.
It is hard to understand such a behavior within a hadron rescattering 
picture unless one assumes that {\em all} relevant chemical reactions 
involve multi-particle channels involving many hadrons. Making such
an assumption clearly pushes the hadronic rescattering model towards 
breakdown because its chemical kinetics would essentially be controlled 
by interactions among clusters of particles involving an unspecifiable 
number of hadrons. {\em It is much more natural to associate this 
kind of behavior with the quark-hadron phase transition} where densely 
spaced and strongly interacting quarks and gluons provide the necessary 
multi-particle clusters, and where the dramatic change in number and 
quality of the effective degrees of freedom within a narrow temperature 
interval generates the dramatic temperature dependence of the chemical 
reaction rates at decoupling which seem to be phenomenologically 
required.  

In such a picture, hadrons are not really well-defined states until after
the quark-hadron phase transition is complete and, at the same time, 
chemical reactions among hadrons have ceased. Hadrons are thus indeed
``born into chemical equilibrium'' \cite{S99} in a process that can be
rightfully called ``statistical hadronization'' 
\cite{Becattini:1997rv,H99,H98}.
If hadrons are formed in this fashion, their measured abundances provide 
a window with a direct view of the QCD quark-hadron phase transition. 

\section{Conclusions}
 
We have shown that the observed impact parameter dependence of the average
temperature and radial flow velocity at {\em kinetic (thermal) freeze-out}
(i.e. at the point where the hadron momentum distributions decouple) can be 
quantitatively understood as a kinetic decoupling process in a 
hydrodynamically expanding source, with freeze-out being driven by the
global expansion of the collision fireball. As Jozs\'o taught us 30 years 
ago \cite{BGZ78}, any such kinetic decoupling
process is controlled by the local competition between temperature 
dependent scattering and hydrodynamic expansion rates, and since the latter
change with impact parameter as a result of the varying initial energy 
density and size of the nuclear collision zone, the resulting average
freeze-out temperature is necessarily impact parameter dependent. The
strength of this impact parameter dependence (i.e. the sensitivity of
the freeze-out temperature to the fireball expansion rate) is inversely
related to the strength of the temperature dependence of the local 
scattering rate. To obtain approximate centrality independence of
the freeze-out temperature, the scattering rate must exhibit an almost 
infinitely steep temperature dependence.  

From this it follows that the observed impact parameter independence of 
the chemical freeze-out temperature in Au+Au collisions at RHIC (i.e. of 
the temperature where the abundances of stable hadron species decouple) 
cannot be consistently described as the result of a kinetic decoupling 
process from inelastic hadronic interactions. To obtain the necessary 
extremely steep temperature dependence of the inelastic scattering rate 
($\sim T^n$ with $n\gtrsim20$) requires that at the freeze-out point 
{\em all} chemical reactions are dominated by multi-hadron interactions 
involving many more than two colliding particles, in which case it seems 
unlikely that one will ever be able to describe this process quantitatively 
in hadronic language. 

In our opinion the only theoretically consistent interpretation of the STAR
data on chemical freeze-out is to associate the steepness of the the
temperature dependence of chemical equilibration rates with a phase
transition (in this case the quark-hadron transition). In this transition
the hadrons are produced statistically and distributed among different 
species according to the principle of maximum entropy, via a multitude of 
complicated microscopic channels involving large numbers of strongly 
interacting quarks and gluons. In this sense the hadrons are ``born into
chemical equilibrium'' in an environment that is too dilute and expands too
rapidly to allow for {\em any} further inelastic reactions among the hadrons.

$T_\mathrm{kin}$ and $T_\mathrm{chem}$ thus stand on conceptually 
different footings. $T_\mathrm{chem}$ is a Lagrange multiplier related 
by the Maximum Entropy Principle to the critical energy density 
$e_c$ for hadronization. Its universality in $e^+e^-$, $pp$, and $AA$ 
collisions of all centralities shows that at $e_c$ a phase transition 
occurs. Hadrons are formed during this transition in a statistical 
process subject to the Principle of Maximum Entropy.

The absence of inelastic {\em hadronic} rescattering processes allows the 
direct measurement of $T_c$ through $T_\mathrm{chem}$ and thus the 
experimental observation of the phase transition. In this context the
question arises which of the different definitions of the critical
temperature $T_c$ from lattice QCD that were studied in \cite{Aoki:2006br}
is most closely related to the chemical freeze-out temperature 
$T_\mathrm{chem}$ extracted from hadron yield data. It seems unlikely
that hadron yields can be considered frozen out before the hadrons have
more or less recovered their full vacuum masses, and this is related
to the restoration of the chiral condensate $\langle\bar\psi\psi\rangle$
to its vacuum value. We therefore suggest that $T_c(\chi_{\bar\psi\psi})$
\cite{Aoki:2006br} should be the LQCD number most closely related to the 
phenomenological value $T_\mathrm{chem}$. This seems to be consistent
with the actual values extracted in \cite{Aoki:2006br} from LQCD and in
\cite{STAR_Tdec} from hadron yields at RHIC (both are between 150 and 
160 MeV).

The increase of the strangeness saturation factor $\gamma_s$ from $e^+e^-$ 
and $pp$ to heavy-ion collisions and from peripheral to central Au+Au 
collisions at RHIC shows that the lifetime of the QGP (and thus the time 
for chemically equilibrating strange with light quarks) is still limited. 
Only for midcentral to central Au+Au collisions $\gamma_s$ has sufficient 
time to saturate. (Qualitatively similar tendencies are seen in Pb+Pb 
collisions at lower SPS energies \cite{BRS03}.) The primary parton 
production process at the beginning of the collision apparently 
suppresses the production of strange quarks, and it also produces 
$s$ and $\bar s$ locally in pairs, thereby generating spatial 
correlations among $s$ and $\bar s$ which ensure strangeness 
conservation {\em locally}. In a grand canonical description such 
correlations induce a strangeness suppression factor $\gamma_s<1$ 
\cite{Bec}. It takes time to diffuse the strange quarks over the
entire fireball volume to decorrelate them and adjust their abundance
to equilibrium values. Larger initial energy densities in central Au+Au
collisions provide more time until the point of hadronization at 
$e_c\simeq0.7$\,GeV/fm$^3$ is reached than peripheral Au+Au or $e^+e^-$
and $pp$ collisions. 

We close by pointing out that our conclusions about the nature and origin of 
$T_\mathrm{chem}$ can be put to a relatively easy experimental test: It is
well known \cite{Cleymans:1998fq,BRS03} that at low SPS and AGS energies, 
where the net baryon density of the matter created in the collision is 
much larger than at RHIC, the measured chemical decoupling temperatures 
are well below generally accepted estimates for the phase transition 
temperature,
$T_\mathrm{chem}<T_c$. In that case the phase transition can not be the 
origin of the observation of chemical equilibrium yields; hadronic chemical 
reactions must be responsible for lowering the chemical freeze-out 
temperature to values significantly below $T_c$. Since the present work 
has shown that the kinetic decoupling of hadronic chemical reaction rates 
is influenced by the fireball expansion rate, which again depends on 
collision centrality, {\em we expect to see impact parameter dependence of 
$T_\mathrm{chem}$ whenever its value is measured to be well below $T_c$}.
This conclusion would also apply to RHIC collisions if lattice QCD would 
eventually converge to $T_c$ values above 190 MeV as proposed in 
\cite{Karsch_Tc}. In this case we would definitely expect $T_\mathrm{chem}$
to depend on collision centrality. We therefore propose a reanalyzis of 
chemical decoupling data at RHIC with higher statistics in order to 
unambiguously settle this question.

It may be possible to reanalyze existing SPS data to confirm or falsify 
our prediction of centrality dependence of $T_\mathrm{chem}$ at these 
energies. If not, this will be a worthwhile point to address within the 
planned low-energy collision program at RHIC. Clarification of this point 
will be of utmost importance for establishing the observed chemical 
decoupling temperature at RHIC as a direct measurement of the critical 
temperature of the quark-hadron phase transition in QCD. 

\section*{Epilogue}

Through his enthusiasm and openness in personal interaction, his 
frequent hospitality, and his brilliant students whom he sent out 
into the world already at early stages of their developing
scientific careers, J\'ozsef Zim\'anyi has had a lasting influence on the 
research of one of us (UH). The present work is a living testimony to 
this, and we therefore dedicate it to his memory.

%%%%%%%%%%%%%%%%%%%%%%%% References %%%%%%%%%%%%%%%%%%%%%%%%%%%%%%%%%%%%%%%%


\begin{thebibliography}{99}
\bibitem{fn2}
  The chemical decoupling temperatures extracted from the measured hadron 
  abundance ratios depend somewhat on the details of the hadron resonance
  gas model employed: Ref.~\cite{STAR_Tchem} gives 
  $T_\mathrm{chem}=163\pm4$\,MeV, $\mu_B=24\pm4$\,MeV for central Au+Au
  collisions whereas in Ref.~\cite{STAR_Tdec} one finds 
  $T_\mathrm{chem}=157\pm6$\,MeV, $\mu_B=22\pm4$\,MeV for the same
  data. We will here use the values from the centrality dependence study 
  presented in Ref.~\cite{STAR_Tdec} which are consistently on the lower
  end of this range.
\end{thebibliography}
\end{document}